\documentclass[reprint,superscriptaddress,amsmath,amssymb,a4paperaps,prl]{revtex4-1}
\usepackage{dcolumn,amssymb,amsmath,amsfonts,graphicx,latexsym,float,color}
\usepackage{epstopdf}
\usepackage[normalem]{ulem}
\begin{document}

\title{Entanglement Induced Interactions in Binary Mixtures}

\author{J. Chen}
\email {jie.chen@physnet.uni-hamburg.de}
\affiliation{Zentrum f\"ur Optische Quantentechnologien, Universit\"at Hamburg, Luruper Chaussee 149, 22761 Hamburg, Germany}
\author{J. M. Schurer}
\email{jschurer@physnet.uni-hamburg.de}
\affiliation{Zentrum f\"ur Optische Quantentechnologien, Universit\"at Hamburg, Luruper Chaussee 149, 22761 Hamburg, Germany}
\affiliation{The Hamburg Centre for Ultrafast Imaging, Universit\"at Hamburg, Luruper Chaussee 149, 22761 Hamburg, Germany}
\author{P. Schmelcher}
\email{pschmelc@physnet.uni-hamburg.de}
\affiliation{Zentrum f\"ur Optische Quantentechnologien, Universit\"at Hamburg, Luruper Chaussee 149, 22761 Hamburg, Germany}
\affiliation{The Hamburg Centre for Ultrafast Imaging, Universit\"at Hamburg, Luruper Chaussee 149, 22761 Hamburg, Germany}
\date{\today}

\begin{abstract}
We establish a conceptual framework for the identification and the characterization of induced interactions in binary mixtures and reveal their intricate relation to entanglement between the components or species of the mixture. Exploiting an expansion in terms of the strength of the entanglement among the two species enables us to deduce an effective single-species description. In this way, we naturally incorporate the mutual feedback of the species and obtain induced interactions for both species which are effectively present among the particles of same type. Importantly, our approach incorporates few-body and inhomogeneous systems extending the scope of induced interactions where two particles interact via a bosonic bath-type environment. Employing the example of a one-dimensional ultracold Bose-Fermi mixture, we obtain induced Bose-Bose and Fermi-Fermi interactions with short-range attraction and long-range repulsion. With this, we show how beyond species mean-field physics visible in the two-body correlation functions can be understood via the induced interactions.
\end{abstract}

\maketitle

\emph{Introduction}.---For a binary mixture, it has become a well-established concept that the exchange of bosonic (quasi-) particles provided by one species, leads to induced interactions effectively present between the particles of the other species. A paradigmatic example is electrons in a crystalline solid which may acquire an attractive interaction by the exchange of phonons, according to the famous Fr\"ohlich Hamiltonian \cite{ind_Frohlich}. Such a concept significantly simplified the way of studying composite systems while, at the same time, it successfully describes the in-depth physics including the celebrated BCS theory for superconductivity \cite{theory_bcs}. Note, however, that the physics of the Fr\"ohlich Hamiltonian focuses on the so-called system-bath regime where the feedback of the electrons on the phonons is usually neglected.

Recent experimental progress in ultracold atomic systems has made them an ideal platform for the investigation on atomic mixtures \cite{mixture_exp_bb_1, mixture_exp_bb_2, mixture_exp_bb_3, mixture_exp_bf_1, mixture_exp_bf_2, mixture_exp_bf_3, mixture_exp_bf_4, mixture_exp_bf_5, mixture_exp_bf_6, mixture_exp_ff_1, mixture_exp_ff_2, mixture_exp_ff_3}. For such mixtures, the induced interaction \cite{ind_bcs} is of particular interest. For example in a Bose-Fermi mixture, the latter can alter the Fermi-Fermi interaction \cite{ind_cold_bf} or the Bose-Bose interaction \cite{ind_cold_bb}. Beside the efforts on macroscopic ensembles, recent experiments \cite{few_exp1, few_exp2, few_exp3, few_exp4, few_exp5} also focus on few-body physics revealing, for example, the formation of the Fermi sea \cite{few_exp1}, the fermionic pairing with an attractive interaction \cite{few_exp2}, as well as the fermionization of two distinguishable fermions \cite{few_exp5}, which pave the way for future experiments on few-body mixtures. Here, however, the notion of an infinitely extended bath-type environment of noninteracting modes ceases to be valid. 

The focus of this work is to establish a conceptual framework for induced interactions. In particular, we show how the interspecies entanglement creates the induced interactions and we derive a general expression for the effective single-species Hamiltonians. Importantly, the analysis of the induced interaction is derived from the eigenstates, which in general can be obtained by any \textit{ab~initio} numerical method, making it applicable for both the ground state and the excited states of the mixture. Moreover, our approach naturally incorporates few-body and inhomogeneous systems extending the scope of induced interactions where two particles interact via a bosonic bath-type environment. 
By using a one-dimensional ultracold Bose-Fermi mixture as an exemplary system, we illustrate how induced interactions can give qualitative insight into a many-body state; e.g., the two-body correlations not captured in the species mean-field can be understood via the induced interactions and the reduction to an effective single-species description is qualitatively valid.

\emph{Theoretical approach}.---The generic Hamiltonian for a binary mixture is given by $\hat{H} =\hat{H}_{A}+\hat{H}_{B}+\hat{H}_{AB}$, where $\hat{H}_{A (B)}$ denotes the Hamiltonian for species $A$ and $B$, respectively. $\hat{H}_{AB}$ represents the interaction between the two species. For the moment, we restrict $\hat{H}_{AB}$ to be a local two-body interaction. Employing a Schmidt decomposition \cite{Schmidt}, an exact eigenstate of the mixture can be uniquely written in the form 
\begin{equation}
|\Psi \rangle = \sum_{i=1}^{\infty} \sqrt{\lambda_{i}} ~|\psi_{i}^{A}\rangle  |\psi_{i}^{B}\rangle, \label{psi}
\end{equation}
where $ \lambda_{i} $ are the Schmidt numbers with $\lambda_1 > \lambda_2 > \cdots$, which are real positive numbers and obey the constraint $\sum_{i} \lambda_i = 1$ originating from the normalization of the wave function $|\Psi \rangle$. The Schmidt numbers directly reveal the strength of the interspecies entanglement, note that for the case $\lambda_{1} = 1$ while all $\lambda_{i \neq 1} = 0$, the mixture is nonentangled \cite{Schmidt}. The $|\psi_{i}^{\sigma}\rangle $ denotes the $i$th Schmidt state for species $\sigma$ with $\sigma = A (B)$. In addition, all Schmidt states $\{|\psi_{i}^{\sigma}\rangle\} $ form an orthonormal basis. It is worthwhile to note that the species mean-field (SMF) approximation assumes the total wave function to be of a simple product form, i.e.~$|\Psi \rangle = |\psi^{A}_{\text{SMF}}\rangle |\psi^{B}_{\text{SMF}}\rangle$, as aforementioned, it excludes entanglement among the two species. For a mixture described by $\hat{H}$, such a product ansatz implies that the mutual impact of the species is merely an additional (induced) potential \cite{Supplemental Material}. It is also worth noting that, the SMF approximation allows for arbitrarily large intraspecies correlations, which is, of course, beyond the mean-field approximations for mixtures  \cite{BEC_Pethick}.

Projecting with the $q$th Schmidt state $ \langle \psi_{q}^{\bar{\sigma}} | $ and multiplying with the $\sqrt{\lambda_{q}}$ from the left onto the time-independent Schr\"odinger equation for the mixture, we obtain
\begin{equation}
\sum_{i=1}^{ \infty} \sqrt{\lambda_{q}} \sqrt{\lambda_{i}} \langle \psi_{q}^{\bar{\sigma}}|\hat{H} |\psi_{i}^{\bar{\sigma}}\rangle  |\psi_{i}^{\sigma}\rangle = \mu_{q} |\psi_{q}^{\sigma}\rangle, \label{eom}
\end{equation}
with $q \in [1,\infty) $, $\bar{\sigma} = B (A)$ for $\sigma = A (B)$, and $\mu_{q} = \lambda_{q} E$, with $E$ being the eigenenergy of the state $|\Psi \rangle$. Moreover, due to the orthogonality of the Schmidt states,  we have  $\mu_{q} = \sum_{j} \sqrt{\lambda_{q}} \sqrt{\lambda_{j}} ~\textit{t}_{qj}$, with $\textit{t}_{qj} = \langle \psi_{q}^{\sigma}| \langle \psi_{q}^{\bar{\sigma}}|\hat{H}|\psi_{j}^{\bar{\sigma}}\rangle |\psi_{j}^{\sigma} \rangle $ representing the transition amplitude between the Schmidt-state products $ |\psi_{j}^{\bar{\sigma}}\rangle |\psi_{j}^{\sigma} \rangle$ and $ |\psi_{q}^{\bar{\sigma}}\rangle |\psi_{q}^{\sigma} \rangle$ \cite{MX, ML-B}. Note that, as long as the Hamiltonian of the mixture possesses the time-reversal symmetry, the eigenfunction $|\Psi \rangle$ can always be chosen real \cite{quantum_mechanics_Landau}, and since all the Schmidt numbers $\sqrt{\lambda_{i}}$ are real as well \cite{Schmidt}, the $\textit{t}_{qj}$ are real numbers, i.e., $\textit{t}_{qj}= \textit{t}_{jq}$.

With this knowledge, we rewrite Eq.~\eqref{eom} as
\begin{align}
&\lambda_{1} \textit{H}_{11}^{\bar{\sigma}} |\psi_{1}^{\sigma}\rangle+ \sum_{i \neq 1} \sqrt{\lambda_{1}}\sqrt{\lambda_{i}} \textit{H}_{1i}^{\bar{\sigma}} |\psi_{i}^{\sigma} \rangle = \mu_{1} |\psi_{1}^{\sigma}\rangle,\label{eom_1}\\
&|\psi_{q}^{\sigma}\rangle = M_{q} \sum_{j \neq q} \sqrt{\lambda_{j}}\sqrt{\lambda_{q}} \textit{H}_{qj}^{\bar{\sigma}} |\psi_{j}^{\sigma} \rangle ~~~~~~ (q > 1), \label{eom_q}
\end{align}
with $\textit{H}_{ij}^{\bar{\sigma}} = \langle \psi_{i}^{\bar{\sigma}}|\hat{H}|\psi_{j}^{\bar{\sigma}}\rangle$ and $M_{q} = \left[ \mu_{q} - \lambda_{q} \textit{H}_{qq}^{\bar{\sigma}} \right]^{-1}$. Substituting Eq.~\eqref{eom_q} into Eq.~\eqref{eom_1} yields the expression  
\begin{equation}
\lambda_{1} \textit{H}_{11}^{\bar{\sigma}} |\psi_{1}^{\sigma}\rangle  + \sum_{i \neq 1} \sum_{j \neq i} \sqrt{\lambda_{1}} \lambda_{i}  \sqrt{\lambda_{j}}  \textit{H}_{1i}^{\bar{\sigma}}  M_{i} \textit{H}_{ij}^{\bar{\sigma}} |\psi_{j}^{\sigma}\rangle  = \mu_{1} |\psi_{1}^{\sigma}\rangle \label{eom_mf_exact}.
\end{equation} 

So far the derivation is completely general and does not include any approximations. However, now we would like to focus on the situation where the two species are weakly entangled. The weak-entanglement regime is defined via 
\begin{equation}
\sqrt{\lambda_{1}} \approx 1 ~~\text{ and} ~~~ \sqrt{\lambda_{i \neq 1}} \ll 1; \label{entangle-condition}
\end{equation}
i.e., the first Schmidt state carries the dominant weight, which explains our focus on $|\psi_{1}^{\sigma}\rangle$ in Eqs.~\eqref{eom_1} and \eqref{eom_mf_exact}. 
Before proceeding, let us briefly explicate the reasons for focusing on the weak-entanglement regime: 
(a) Focusing on the weak interspecies interaction regime, in general, only weak entanglement can be expected. Hence, also experimentally the weak-entanglement regime can be achieved in ultracold atomic systems with the aid of Feshbach or confinement induced resonances \cite{feshbach_resonance, a_1d, weak_interaction}. Moreover, we witness that the validity of the weak-entanglement regime extends far beyond the perturbative regime (see below). 
(b) It is also permissible to mitigate the interspecies entanglement by using a unitary transformation of the Hamiltonian, such as the Fr\"ohlich-Nakajima transformation or the Lee-Low-Pines transformation for polarons \cite {unitary_transformations}.
(c) In the weak-entanglement regime, the first Schmidt state contains the leading contribution to the properties of the many-body state, which suggests the possibility for pure state analysis. This can be seen through the decomposition of the reduced density matrix for the $\sigma$ species,
\begin{align}
\rho_{\sigma} &= \text{tr}_{\bar{\sigma}} |\Psi \rangle \langle \Psi|  =\sum_{i} \lambda_{i}  |\psi_{i}^{\sigma} \rangle \langle \psi_{i}^{\sigma} | \approx  |\psi_{1}^{\sigma} \rangle \langle \psi_{1}^{\sigma} |, \label{dmat-n decomposition}
\end{align}
here $ \text{tr}_{\bar{\sigma}} $ denotes tracing out the $\bar{\sigma}$ species. Besides, we employed the conditions \eqref{entangle-condition} in the last step, assuming all the terms of order $\lambda_{i \neq 1}$ are negligible. 

Equipped with this knowledge, we perform a Taylor expansion on Eq.~\eqref{eom_mf_exact}. This is achieved by assuming all $\sqrt{\lambda_{i \neq 1}}$ to be of order $\delta$ (or smaller) with $\delta \ll 1$ and neglecting the terms of O($\delta^2$). As our main result, we obtain the effective Hamiltonian for species $\sigma$, which is given by
\begin{equation}
\hat{H}_{\text{eff}}^{\sigma} = \textit{H}_{11}^{\bar{\sigma}} + \sum_{i \neq 1} \frac{ \sqrt{\lambda_{i}} \textit{H}_{1i}^{\bar{\sigma}} \textit{H}_{i1}^{\bar{\sigma}}}{\textit{t}_{1i}}, \label{effctive_Hamiltonian} 
\end{equation}
with the associated effective Schr\"odinger equation
\begin{equation}
\hat{H}_{\text{eff}}^{\sigma} |\psi_{\text{eff}}^{\sigma}\rangle  = E_{\text{eff}}^{\sigma} |\psi_{\text{eff}}^{\sigma}\rangle. \label{eom_mf_1st_order}
\end{equation}
Equation \eqref{eom_mf_1st_order} not only introduces a significant simplification in the study of mixtures but also allows for profound insights. Indeed it is in many cases extremely difficult to gain deep insights and extract relevant mechanisms in coupled binary mixtures even if the complete many-body wave function is attainable. Here the analytical and interpretational power of Eq.~\eqref{eom_mf_1st_order} comes into play. In order to explain this in more detail, the following considerations are in order.
(i) The effective state $|\psi_{\text{eff}}^{\sigma}\rangle$ is the eigenstate of $\hat{H}_{\text{eff}}^{\sigma}$ whose eigenvalue is closest to $E_{1}^{\sigma} =  \langle \psi_{1}^{\sigma} | \hat{H}_{\text{eff}}^{\sigma} |\psi_{1}^{\sigma}\rangle $ \cite{Supplemental Material}. Importantly, $|\psi_{\text{eff}}^{\sigma}\rangle$ can be understood as an approximation to $ |\psi_{1}^{\sigma}\rangle$ in Eq.~\eqref{eom_mf_exact}, which contains the dominant physics in the spirit of Eq.~\eqref{dmat-n decomposition}.
(ii) It is worth noting that the effective Hamiltonian $\hat{H}_{\text{eff}}^{\sigma} $ depends on the many-body state of the mixture. Importantly, a state-dependent Hamiltonian indeed reflects the fact that the nature of the mixture may differ significantly for different parameters, in particular, for few-particle systems. Since $|\Psi \rangle $ is a general eigenstate of the mixture, the deduction of the effective Hamiltonian $\hat{H}_{\text{eff}}^{\sigma}$ is applicable for both the ground state and the excited states of the mixture. 
(iii) If the mixture is nonentangled, the exact eigenfunction $|\Psi\rangle $ is of a product form, i.e., $\sqrt{\lambda_i} = 0$ for all $i>1$; therefore, only the first term in the effective Hamiltonian \eqref{effctive_Hamiltonian} is present, which corresponds to the SMF case. Since $\hat{H}_{AB}$ is a local two-body interaction, the effective Hamiltonian becomes $\hat{H}_{\text{eff}}^{\sigma} = \hat{H}_{\sigma} + \hat{V}^{\sigma}_{\text{SMF}}$ with $\hat{V}^{\sigma}_{\text{SMF}}$ being an additional (induced) potential. We will refer to $\hat{V}^{\sigma}_{\text{SMF}}$ as the SMF induced potential in the following discussions. Mathematically, it is the partial trace with respect to the species $\bar{\sigma}$ over the interspecies interaction $\hat{H}_{AB}$ \cite{Supplemental Material}. 
(iv) For the weak-entanglement regime, we obtain the physics beyond the SMF approximation. Since $\sqrt{\lambda_i} \ll 1$ for $i>1$, the first term on the right-hand side of Eq.~\eqref{effctive_Hamiltonian} dominates the effective Hamiltonian and is reminiscent of the above SMF effective Hamiltonian. The second term, which solely originates from the interspecies entanglement contains, besides additional potential term (see below), the induced interaction $[\propto (H^{\bar{\sigma}}_{1i})^{2}]$. Note that, the induced interactions always exist in both species (see also below). Importantly, this induced interaction is mediated via the Schmidt states $|\psi_{i \neq 1}^{\bar{\sigma}}\rangle$ from species $\bar{\sigma}$. Moreover, it is a series with monotonically decreasing prefactors $\sqrt{\lambda_{i}}$, which suggests the possibility to truncate the sum over $i$, since the contributions from the Schmidt states with large $i$ can be neglected \cite{Supplemental Material}. 
(v) The last important observation is that for the case $N_{\sigma} \gg N_{\bar\sigma}$, the so-called system-bath regime, the induced interaction in the bath species $\sigma$ becomes negligible. This is because $ \textit{H}_{1i}^{\bar{\sigma}} \textit{H}_{i1}^{\bar{\sigma}} $ is proportional to $(N_{\bar{\sigma}})^{2}$, while $\textit{t}_{1i} = \langle \psi_{1}^{\sigma}| \langle \psi_{1}^{\bar{\sigma}}|\hat{H}|\psi_{i}^{\bar{\sigma}}\rangle |\psi_{i}^{\sigma} \rangle $ scales with $N_{\sigma}N_{\bar{\sigma}}$, leading to an induced interaction for species $\sigma$ directly proportions to $N_{\bar{\sigma}} / N_{\sigma}$. Hence, for $N_{\sigma} \gg N_{\bar\sigma}$, there is negligible induced interaction among the $\sigma$ particles, the $\sigma$ species thereby becomes an ideal bath-type system. Meanwhile, the induced interaction in the $\bar {\sigma}$ species becomes increasingly important.

\emph{Induced interactions in a Bose-Fermi mixture}.---Let us now elucidate the induced interactions and induced potentials for a mixture which is not a typical example of the applications of induced interactions: we discuss the few-body ensemble of a 1D  ultracold Bose-Fermi mixture. The model Hamiltonian (in harmonic units) is $\hat{H} = \hat{H}_{b}+\hat{H}_{f}+\hat{H}_{bf}$, where
\begin{align}
\hat{H}_{\sigma}&=\int dx~\hat{\psi}^{\dagger}_{\sigma}(x) \textit{h}_{\sigma}(x) \hat{\psi}_{\sigma}(x), ~~~ (\sigma = b,f) \nonumber\\
\hat{H}_{bf} &= {g_{bf}}\int dx~\hat{\psi}^{\dagger}_{f}(x) \hat{\psi}^{\dagger}_{b}(x) \hat{\psi}_{b}(x) \hat{\psi}_{f}(x), \label{Hamiltonian_bf}
\end{align}
with $\textit{h}_{\sigma}(x) = -\frac{1}{2}\frac{\partial^{2}}{\partial x^{2}}+\frac{1}{2}x^{2}$ being the single-particle Hamiltonian with harmonic confinement and $\hat{H}_{bf}$ containing a contact interaction \cite{delta_interaction, a_1d} between the two species with interaction strength $g_{bf}$. To demonstrate the physics originating from the interspecies entanglement, we focus on the case of interspecies interaction only with $g_{bf} = 1$ and explore a mixture with two fermions and two bosons $N_{f} = N_{b}=2$. Note that our approach is also valid when Bose-Bose and Fermi-Fermi interactions are taken into account. Moreover, this interaction strength is far beyond the perturbative regime with the interaction energy $E_{\text{int}} = 0.82 $ being comparable to the total kinetic energy $E_{k} = 1.38 $. For such a Bose-Fermi mixture, the associated effective Hamiltonian for species $\sigma$ is
\begin{equation}
\hat{H}_{\text{eff}}^{\sigma}  = \hat{H}_{\sigma} + \hat{V}_{\text{ind}}^{\sigma} + \hat{H}_{\text{ind}}^{\sigma}, \label{eom_bf_1st_order}
\end{equation}
with
\begin{align}
\hat{V}_{\text{ind}}^{\sigma} & = \int dx \left[ \textit{V}_{1}^{\sigma}(x) + \textit{V}_{\text{no}}^{\sigma}(x) \right]  \hat{\psi}^{\dagger}_{\sigma}(x) \hat{\psi}_{\sigma}(x) \label{v_eff_bf},  \\
\hat{H}_{\text{ind}}^{\sigma} &=  \frac{1}{2}\int dx_{1} dx_{2}~ \textit{H}_{\text{ind}}^{\sigma} (x_{1}, x_{2}) \hat{\psi}^{\dagger}_{\sigma}(x_{1}) \hat{\psi}^{\dagger}_{\sigma}(x_{2}) \hat{\psi}_{\sigma}(x_{2}) \hat{\psi}_{\sigma}(x_{1}),\label{induced_interaction_bf}
\end{align}
representing the induced potential and induced interaction, respectively (see also below). In the following discussions, both the induced interactions and the induced potentials are obtained from \textit{ab~initio} ML-MCTDHX simulations \cite{MX}. In addition, we also compare to the results of exact diagonalization (ED) simulations, showing a qualitative agreement \cite{Supplemental Material}.

\begin{figure}[H]
  \centering
  \includegraphics[width=0.5\textwidth]{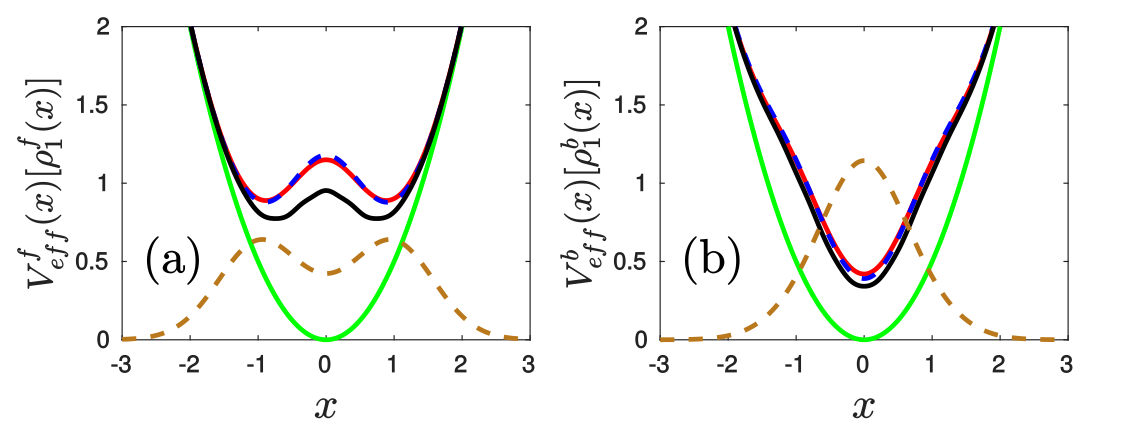}\hfill
  \caption{Effective potentials and reduced one-body density. The effective potentials for the fermionic (a) and the bosonic (b) species for the noninteracting case ($\frac{1}{2}x^{2}$) (green solid lines), the SMF case [$\frac{1}{2}x^{2} + \textit{V}_{\text{SMF}}^{\sigma}(x)$](blue dashed lines), the case $[\frac{1}{2}x^{2} + \textit{V}_{1}^{\sigma}(x)]$ (red solid lines), and beyond the SMF [$\frac{1}{2}x^{2} + \textit{V}_{\text{ind}}^{\sigma}(x)$](black solid lines). In addition, the reduced one-body densities for the fermionic (a) and the bosonic (b) species (brown dashed lines) are depicted as well.}
\label{effective_potentials}
\end{figure}

The induced potential consists of two terms with the first one,
\begin{equation}
\textit{V}_{1}^{\sigma} (x)= g_{bf}  \gamma_{11}^{\bar{\sigma}} (x), \label{v_smf}
\end{equation}
which can be viewed as the SMF contribution. Here, $\gamma_{ij}^{\bar{\sigma}}(x) = \langle\psi_{i}^{\bar{\sigma}}|\hat{\psi}^{\dagger}_{\bar{\sigma}} (x) \hat{\psi}_{\bar{\sigma}}(x) |\psi_{j}^{\bar{\sigma}}\rangle $ is the reduced one-body transition matrix element for species $\bar{\sigma}$ \cite{MX}. From Eq.~\eqref{v_smf}, we see that $\textit{V}_{1}^{\sigma}(x)$ is proportional to both interspecies interaction strength $g_{bf}$ and $\gamma_{11}^{\bar{\sigma}}$, the contribution made by the first Schmidt state to the reduced one-body density $\rho^{\bar{\sigma}}_{1}$ for species $\bar{\sigma}$. As discussed above, in the weak-entanglement regime we have $\textit{V}_{1}^{\sigma}(x) \approx  V_{\text{SMF}}^{\sigma}(x) $; i.e., $\textit{V}_{1}^{\sigma}(x) $ highly resembles the SMF induced potential [cf.~Figs.~\ref{effective_potentials}(a) and \ref{effective_potentials}(b), blue dashed and red solid lines]. The second term $\textit{V}_{\text{no}}^{\sigma}(x)$ is
\begin{equation}
\textit{V}_{\text{no}}^{\sigma} (x) = g_{bf} \sum_{i \neq 1} \frac{ \sqrt{\lambda_{i}}}{\tilde{t}_{1i}} \left[ \gamma_{1i}^{\bar{\sigma}}(x) \gamma_{i1}^{\bar{\sigma}} (x) + 2 \beta_{1i}^{\bar{\sigma}} \gamma_{i1}^{\bar{\sigma}}(x) \right], \label{v_ind}
\end{equation}
with $\tilde{t}_{1i}  =  \int dx~ \gamma_{1i}^{\sigma}(x) \gamma_{1i}^{\bar{\sigma}}(x) $ and $\beta_{1i}^{\bar{\sigma}} = \langle\psi_{1}^{\bar{\sigma}}|\hat{H}_{\bar{\sigma}} |\psi_{i}^{\bar{\sigma}}\rangle$, in which $\gamma_{1i}^{\bar{\sigma}} \gamma_{i1}^{\bar{\sigma}}$ results from normal ordering of $\langle \psi_{1}^{\bar{\sigma}}|\hat{H}_{\sigma \bar{\sigma} }|\psi_{i}^{\bar{\sigma}}\rangle \langle \psi_{i}^{\bar{\sigma}}|\hat{H}_{\sigma \bar{\sigma}} |\psi_{1}^{\bar{\sigma}}\rangle $ in Eq.~\eqref{effctive_Hamiltonian} while $\beta_{1i}^{\bar{\sigma}} \gamma_{i1}^{\bar{\sigma}} $ stems from cross terms such as $\langle \psi_{1}^{\bar{\sigma}}|\hat{H}_{\bar{\sigma}}|\psi_{i}^{\bar{\sigma}}\rangle \langle \psi_{i}^{\bar{\sigma}}|\hat{H}_{\sigma \bar{\sigma}}|\psi_{1}^{\bar{\sigma}}\rangle$. It is fair to treat $\textit{V}_{\text{no}}^{\sigma}(x)$ as a correction to $ \textit{V}_{1}^{\sigma}(x) $, due to the prefactors $\sqrt{\lambda_{i}}$, rendering the magnitude of $\textit{V}_{\text{no}}^{\sigma}(x)$ small in comparison to $ \textit{V}_{1}^{\sigma}(x)$. In order to elaborate on the net confinement that a particle feels, we introduce the effective potentials as $\textit{V}_{\text{eff}}^{\sigma}(x) = \frac{1}{2}x^{2} + \textit{V}_{\text{ind}}^{\sigma}(x) $, as shown in Fig.~\ref{effective_potentials}(a) and \ref{effective_potentials}(b) (black solid lines). We observe that, due to the presence of induced potentials, the effective potentials for both species deviate significantly from the original harmonic confinement forming either a double-well pattern (fermionic) or a tighter confinement (bosonic). In addition, the SMF effective potentials are presented as well [cf.~Figs.~\ref{effective_potentials}(a) and \ref{effective_potentials}(b), blue dashed lines].

\begin{figure}[htp]
\centering
\includegraphics[width=0.5\textwidth] {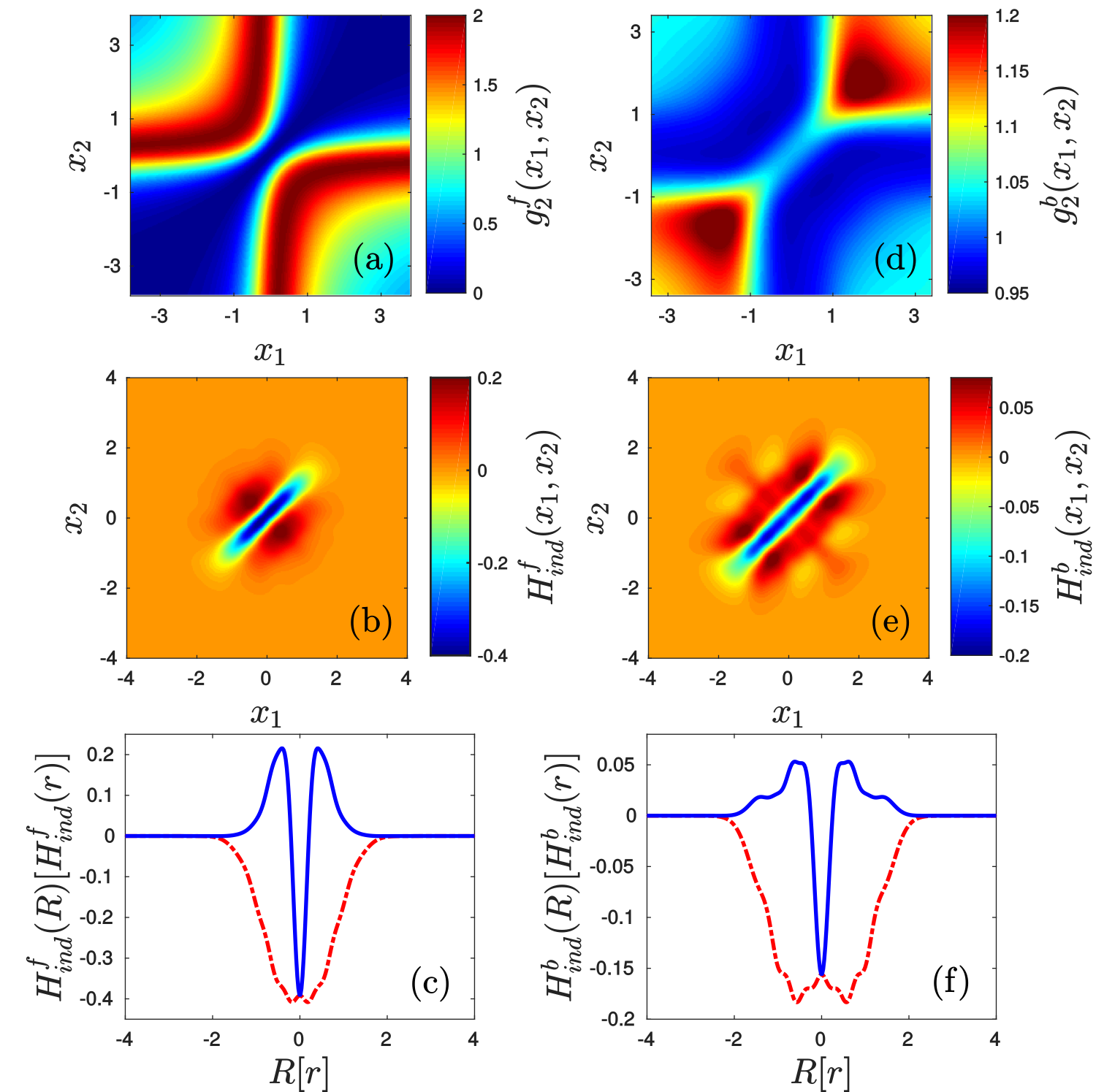} \hfill
\caption{Induced interactions and pair-correlation functions. The upper panels are pair-correlation functions $g^{\sigma}_2(x_1,x_2)$ for the fermionic (a) and the bosonic (d) species. The middle and lower panels are the induced interactions $\textit{H}_{\text{ind}}^{\sigma}(x_1,x_2)$ among the fermions (b),(c) and the bosons (e),(f) together with its diagonals [red dashed lines, $x_1=x_2$ and $R = (x_1+x_2)/2$] and off diagonals (blue solid lines, $x_1=-x_2$ and $r = x_1-x_2$).} 
\label{induced_interactions} 
\end{figure}

Now we turn to the induced interaction, which reads 
\begin{equation}
\textit{H}_{\text{ind}}^{\sigma}(x_{1}, x_{2}) = g_{bf}  \sum_{i \neq 1} \frac{ 2\sqrt{\lambda_{i}}}{\tilde{t}_{1i}} \gamma_{1i}^{\bar{\sigma}}(x_{1}) \gamma_{i1}^{\bar{\sigma}}(x_{2}). \label {ind-int_bf}
\end{equation}
Unlike the Fr\"ohlich Hamiltonian where the induced interaction is a second order perturbation term with respect to $g_{bf}$, here, it derives from the first order expansion with respect to the square root of the Schmidt numbers $\sqrt{\lambda_{i \neq 1}}$; therefore, it does not restrict the interaction strength $g_{bf}$ to be small. In Figs.~\ref{induced_interactions}(b) and \ref{induced_interactions}(e) we present the induced interactions among the fermions and the bosons, respectively. Importantly, the computed induced interaction preserves the particle exchange symmetry $ \textit{H}_{\text{ind}}^{\sigma}(x_{1}, x_{2}) = \textit{H}_{\text{ind}}^{\sigma}(x_{2}, x_{1}) $ for identical particles as well as the parity symmetry $ \textit{H}_{\text{ind}}^{\sigma}(x_{1}, x_{2}) = \textit{H}_{\text{ind}}^{\sigma}( -x_{1}, -x_{2}) $ of the original Hamiltonian [cf. Eq.~\eqref{Hamiltonian_bf}].
Furthermore, we notice that, unlike the contact Bose-Fermi interaction, the induced interaction is long ranged and becomes, depending on the relative coordinate $r = x_{1} - x_{2}$, attractive for small particle distances, repulsive for increasing $r$, and vanishes at large $r$ [cf.~Figs.~\ref{induced_interactions}(c) and \ref{induced_interactions}(f)]. Apart from the dependence on the relative coordinate $r$, the induced interaction also varies as a function of the mean position of the particles $R = (x_{1} + x_{2}) /2$, since our system is not translationally invariant. These novel features differ from the situation in homogenous systems, where only the relative coordinate is involved and is usually attractive \cite{ind_cold_bf, ind_cold_bb, BEC_Pethick}. In addition, albeit the similar spatial patterns, the induced interactions for the bosons and the fermions show noticeable differences with respect to their strengths and ranges. We observe that the fermionic induced interaction has almost a twice as large maximal value as the bosonic induced interaction, while its range is smaller.

\begin{figure}[htp]
  \centering
  \includegraphics[width=0.5\textwidth]{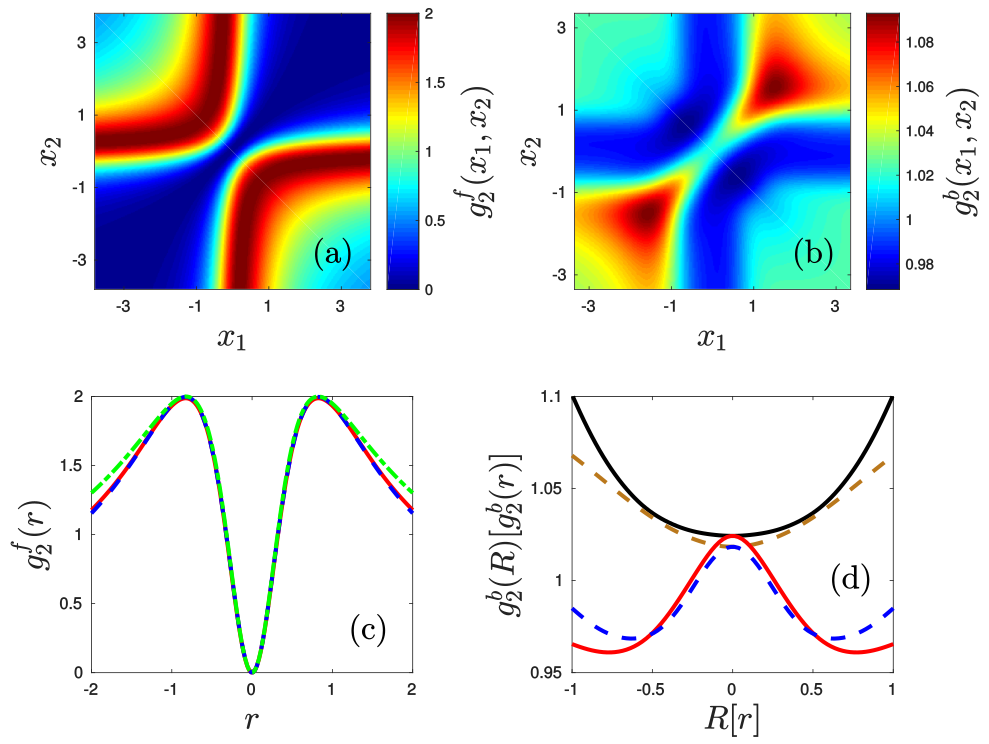}\hfill
 \caption{Comparisons of pair-correlation functions. The pair-correlation functions $g^{\sigma}_2(x_1,x_2)$ computed by the effective Hamiltonian $\textit{H}_{\text{eff}}^{\sigma}$ for fermionic (a),(c) and the bosonic (b),(d) species, together with its off diagonals (blue dashed lines) and diagonals (brown dashed lines). Note that, the diagonal of $g^{f}_{2}$ is always zero due to the Pauli principle. For comparison, the off diagonal of $g^{f}_{2}$ for the SMF (green dash-dot line) and the ML-MCTDHX simulation (all solid lines) are depicted as well.}
\label{g2_comparison}
\end{figure}

The induced interaction can be an essential quantity for understanding in depth the physics of the mixture. In particular, it can explain the behavior of the pair-correlation function \cite{g2}, which is given by
\begin{equation}
g_{2}^{\sigma}(x_{1}, x_{2})  = \frac{\rho^{\sigma}_{2} (x_{1},x_{2})}{\rho^{\sigma}_{1} (x_{1}) \rho^{\sigma}_{1} (x_{2})}, \label{g2}
\end{equation}
with $\rho^{\sigma}_{2} (x_{1},x_{2})$ and $\rho^{\sigma}_{1} (x) $ being the reduced two- and one-body density, respectively. Physically, $\rho^{\sigma}_{2} (x_{1},x_{2})$ denotes a measure for the probability of finding one particle at $x_1$ while the second is at $x_2$. Through the division by the one-body densities, the $g^{\sigma}_{2}$ function excludes the impact of the inhomogeneous density distribution and thereby directly reveals the spatial two-particle correlations induced by the interaction. In Figs.~\ref{induced_interactions}(a) and \ref{induced_interactions}(d) we present the $g^{\sigma}_{2}$ functions obtained from the \textit{ab~initio} ML-MCTDHX simulations. We find that, due to the interspecies entanglement, the bosonic pair-correlation function deviates significantly from the result of the SMF approximation given by $g^{b}_2(x_{1}, x_{2}) = 1$ for all $x_{1}$ and $x_{2}$. Furthermore, based on the profile of the bosonic induced interaction [cf.~Fig.~\ref{induced_interactions}(e)], we can conclude that the attractive part of the interaction increases the probability of finding two bosons next to each other and, hence, results in $g_{2}^{b} > 1$ near the diagonal ($x_1 = x_2$), while the repulsive part suppresses bosonic bunching and leads to $g_{2}^{b}<1$ at larger relative distance [cf.~Fig.~\ref{induced_interactions}(d)]. In contrast, the induced interaction among the fermions has minor impact due to the Pauli-exclusion principle resulting in a $g_{2}^{f}$  similar to the SMF approximation [cf.~Fig.~\ref{induced_interactions}(a) and Figs.~\ref{g2_comparison}(a) and \ref{g2_comparison}(c)].

In order to test the applicability of our effective Hamiltonian, we compute the pair-correlation functions for both species from the ground state of the effective single-species Hamiltonian \eqref{eom_bf_1st_order} and present them in the Fig.~\ref{g2_comparison}. Compared to the exact $g_{2}^{\sigma}$ functions, we find qualitative agreement between both methods suggesting that our effective Hamiltonian $\textit {H}_{\text{eff}}^{\sigma}$ can evidently explain the important physics. 

\emph{Conclusions}.---We have introduced a new approach to derive the induced interactions for a binary mixture in the weak-entanglement regime. A theoretical framework leading to an effective single-species Hamiltonian which contains the induced interaction as well as an induced potential. Importantly, the induced interaction directly originates from the physics beyond the species mean-field and unveils the relationship to the interspecies entanglement. As an example, we employ a 1D ultracold Bose-Fermi mixture to explore our scheme. The profile of the induced interaction for both species contains a short-range attraction and a long-range repulsion which is position dependent due to the inhomogeneity of the system. Finally, we have compared the pair-correlation functions from ML-MCTDHX simulations with simulations of the effective Hamiltonian and find them in good agreement. Our approach opens up the way to study induced interactions in binary mixtures in particular for few-body and inhomogeneous systems. Thereby, it is of specific interest how the functional form of the induced interaction depends on systems properties. Apart from being readily generalizable to the multispecies mixtures via a suitable bipartition, our approach should be extendable to time-dependent entangled states in the future. This generalization would enable us to compare to the commonly employed time-dependent perturbation theory picture. 

\begin{acknowledgments}
The authors acknowledge fruitful discussions with Sven Kr\"onke, Kevin Keiler, and Maxim Pyzh. Moreover, the authors thank Richard Schmidt for a detailed feedback on the manuscript. J.C. and P.S. gratefully acknowledge financial support by the Deutsche Forschungsgemeinschaft (DFG) in the framework of the SFB 925 ``Light induced dynamics and control of correlated quantum systems." The excellence cluster ``The Hamburg Centre for Ultrafast Imaging-Structure: Dynamics and Control of Matter at the Atomic Scale" is acknowledged for financial support.
\end{acknowledgments}


\begin{thebibliography}{10}
\bibitem {ind_Frohlich} H. Fr\"ohlich, Phys. Rev. {\bf 79}, 845 (1950).
\bibitem {theory_bcs} J. Bardeen, L. N. Cooper, and J. R. Schrieffer, Phys. Rev. {\bf 108}, 1175 (1957).

\bibitem{mixture_exp_bb_1} S. B. Papp, J. M. Pino, and C. E. Wieman, Phys. Rev. Lett. {\bf 101}, 040402 (2008).
\bibitem{mixture_exp_bb_2} S. Tojo, Y. Taguchi, Y. Masuyama, T. Hayashi, H. Saito, and T. Hirano, Phys. Rev. A {\bf 82}, 033609 (2010).
\bibitem{mixture_exp_bb_3} D. J. McCarron, H. W. Cho, D. L. Jenkin, M. P. K\"oppinger, and S. L. Cornish, Phys. Rev. A {\bf84}, 011603(R) (2011).
\bibitem{mixture_exp_bf_1}  A. G. Truscott, K. E. Strecker, W. I. McAlexander, G. B. Partridge, and R. G. Hulet, Science {\bf291}, 2570 (2001).
\bibitem{mixture_exp_bf_2}  F. Schreck, L. Khaykovich, K. L. Corwin, G. Ferrari, T. Bourdel, J. Cubizolles, and C. Salomon, Phys. Rev. Lett. {\bf 87}, 080403 (2001). 
\bibitem{mixture_exp_bf_3} Z. Hadzibabic, C. A. Stan, K. Dieckmann, S. Gupta, M. W. Zwierlein, A. G\"orlitz, and W. Ketterle, Phys. Rev. Lett. {\bf88}, 160401 (2002).
\bibitem{mixture_exp_bf_4} G. Roati, F. Riboli, G. Modugno, and M. Inguscio, Phys. Rev. Lett. {\bf 89}, 150403 (2002).
\bibitem{mixture_exp_bf_5} G. Modugno, G. Roati, F. Riboli, F. Ferlaino, R. J. Brecha, and M. Inguscio, Science {\bf 297}, 2240 (2002).
\bibitem{mixture_exp_bf_6} K. G\"unter, T. St\"oferle, H. Moritz, M. K\"ohl, and T. Esslinger, Phys. Rev. Lett. {\bf 96}, 180402 (2006).
\bibitem{mixture_exp_ff_1} M. W. Zwierlein, A. Schirotzek, C. H. Schunck, and W. Ketterle, Science {\bf311}, 492 (2006).
\bibitem{mixture_exp_ff_2} B. DeMarco and D. S. Jin, Science {\bf285}, 1703 (1999).
\bibitem{mixture_exp_ff_3} K. M. O'Hara, S. L. Hemmer, M. E. Gehm, S. R. Granade, and J. E. Thomas, Science {\bf298}, 2179 (2002).
\bibitem {ind_bcs} J. Bardeen, G. Baym, and D. Pines, Phys. Rev. {\bf156}, 207 (1967).
\bibitem {ind_cold_bf} M. J. Bijlsma, B. A. Heringa, and H. T. C. Stoof, Phys. Rev. A {\bf 61}, 053601 (2000). 
\bibitem {ind_cold_bb} J. J. Kinnunen, G. M. Bruun, Phys. Rev. A {\bf 91}, 041605(R) (2015).

\bibitem{few_exp1} A. N. Wenz, G. Z\"urn, S. Murmann, I. Brouzos, T. Lompe, S. Jochim, Science {\bf342}, 457 (2013).
\bibitem{few_exp2} G. Z\"urn, A. N. Wenz, S. Murmann, A. Bergschneider, T. Lompe, and S. Jochim, Phys. Rev. Lett. {\bf111}, 175302 (2013).
\bibitem{few_exp3} S. Murmann, A. Bergschneider, V. M. Klinkhamer, G. Z\"urn, T. Lompe, S. Jochim, Phys. Rev. Lett. {\bf 114}, 080402 (2015).
\bibitem{few_exp4} F. Serwane, G. Z\"urn, T. Lompe, T. B. Ottenstein, A. N. Wenz, and S. Jochim, Science {\bf 332}, 336 (2011).
\bibitem{few_exp5} G. Z\"urn, F. Serwane, T. Lompe, A. N. Wenz, M. G. Ries, J. E. Bohn, and S. Jochim, Phys. Rev. Lett. {\bf 108}, 075303 (2012).
\bibitem {Schmidt}  A. Pathak, \textit{ Elements of Quantum Computation and Quantum Communication}, (Taylor \& Francis, 2013).
\bibitem {Supplemental Material} See Supplemental Material for details on the species mean-field (SMF) approximation, the computational method, the decomposition of the induced interactions, the discussions on convergence, the effective state and the effects from small perturbations, which includes Refs. \cite{MX,ML-B,improved_relax}.

\bibitem {MX} L. Cao, V. Bolsinger, S. I. Mistakidis, G. M. Koutentakis, S. Kr\"onke, J. M. Schurer and P. Schmelcher, J. Chem. Phys. {\bf 147}, 044106  (2017).
\bibitem {ML-B} L. Cao, S. Kr\"onke, O. Vendrell, and P. Schmelcher, J. Chem. Phys. {\bf 139}, 134103 (2013); S. Kr\"onke, L. Cao, O. Vendrell, and P. Schmelcher, New J. Phys. {\bf 15}, 063018 (2013).
\bibitem {improved_relax} H.-D. Meyer, U. Manthe, and L. Cederbaum, Chem. Phys. Lett. {\bf165}, 73 (1990).
\bibitem {BEC_Pethick} C. J. Pethick and H. Smith, \textit{Bose-Einstein Condensation in Dilute Gases}, (Cambridge University Press, New York, 2008).

\bibitem {quantum_mechanics_Landau} L. D. Landau and E. M. Lifshitz, \textit{Quantum Mechanics: non-relativistic theory}, (Pergamon, Oxford, 1977).
\bibitem {feshbach_resonance} C. Chin, R. Grimm, P. Julienne, and E.Tiesinga, Rev. Mod. Phys. {\bf82}, 1225 (2010).
\bibitem {a_1d} M. Olshanii, Phys. Rev. Lett. {\bf 81}, 938 (1998).
\bibitem {weak_interaction} X. Du, Y. Zhang, J. Petricka, and J. E. Thomas, Phys. Rev. Lett. {\bf103}, 010401 (2009).
\bibitem {unitary_transformations} G. D. Mahan, \textit{Many-Particle Physics}, (Springer New York, 2000); W. Nolting,  \textit{Fundamentals of Many-body Physics, Principles and Methods}, (Springer-Verlag Berlin Heidelberg, 2009)
\bibitem {delta_interaction} K. Huang and C. N. Yang, Phys. Rev. {\bf105}, 767 (1957).
\bibitem {g2} K. Sakmann, A. I. Streltsov, O. E. Alon, and L. S. Cederbaum, Phys. Rev. A {\bf78}, 023615 (2008); K. V. Kheruntsyan, D. M. Gangardt, P. D. Drummond, and G. V. Shlyapnikov, Phys. Rev. Lett. {\bf 91}, 040403 (2003).
\end{thebibliography}
\end{document}